# Clonal-Based Cellular Automata in Bioinformatics

*Pokkuluri Kiran Sree[1#], Inampudi Ramesh Babu[2]*

*[1]Department of CSE, JNTU Hyderabad,*

*[2]Department of CSE, Acharya Nagarjuna University, Guntur, India*

**Abstract**- *This paper aims at providing a survey on the problems that can be easily addressed by clonal-based cellular automata in bioinformatics. Researchers try to address the problems in bioinformatics independent of each problem. None of the researchers has tried to relate the major problems in bioinformatics and find a solution using common frame work. We tried to find various problems in bioinformatics which can be addressed easily by clonal based cellular automata. Extensive literature survey is conducted. We have considered some papers in various journals and conferences for conduct of our research. This paper provides intuition towards relating various problems in bioinformatics logically and tries to attain a common frame work with respect to clonal based cellular automata classifier for addressing the same.*

[#]**Corresponding Author E-mail**: *profkiran@yahoo.com*

## 1. INTRODUCTION

The survey mainly identifies various problems in bioinformatics that can be solved easily by clonal-based cellular automata CA [7, 18] classifier like protein coding identification, promoter prediction, protein structure prediction and RNA structure prediction. In Secs.2 and 3 of this paper, clonal-based cellular automata is introduced with AIS-MACA (artificial immune system-based multiple attractor cellular automata). The next Secs.4and 5 of the paper introduces the survey on protein coding region and promoter region prediction. Then the survey on protein structure prediction and RNA structure predictions presented in Sec.6. Section7 presents a common frame work tool output based on clonal CA.

## 2. Types of Cellular Automata

### 2.1 Linear/Additive CA

Linear/Additive CA is a unique class of CA whose characterization is simple as it can be implemented easily by algebraic methods. The transitions between the cells follow the operations of Galosis field (GF). One of the important variations of linear CA is referred to as GF (2)where each cell in the lattice can either be 0 or 1. It is very difficult to produce hybrid CA as it was demanded in many cases using this class.

### 2.2 Group CA

Group CA is used to generate pseudo-random patterns. Serra *et al.*[19]worked on finding the maximum length of the group CA with non-zero states within a single cycle. Rules 90 and 150 are used to formulate the group CA. Since group CA consists of cycles, it cannot be used for pattern classification.

### 2.3 Non-Group CA

The applications of non-group CA are increasing in wider domains, due to the innate potential of classification. The proposed algorithms of this research also fall into this group. Some important classes of this type are studied as single attractor cellular automata (SACA) and multiple attractor cellular automata (MACA), with depth-1 have been used for various applications in designing, authentication and hashing.

## 3. Clonal-based cellular automata

### 3.1 Cellular Automata

Cellular automata consist of a grid of cells with a finite number of states. CA is a computing model which provides a good platform for performing complex computations with the available local information.

CA is defined as four tuple $< G, Z, N, F >$

where, G => Grid (set of cells)

Z => Set of possible cell states

N => Set which describe cells neighborhoods

F => Transition function (rules of automata)

The complex nature of CA arises from bottom-up manner from the spatially extended grid whose interactions are locally defined. A programmer can specify local rules for interaction among neighbors and can study the consequences of the rules with respect to fitness. The selection of the rules which can be applied to solve a particular problem is challenging in CA.

We have implemented MACA [11, 14]which uses fuzzy logic with the help of clonal algorithm. The design of clonal CA is given in Figure 0.

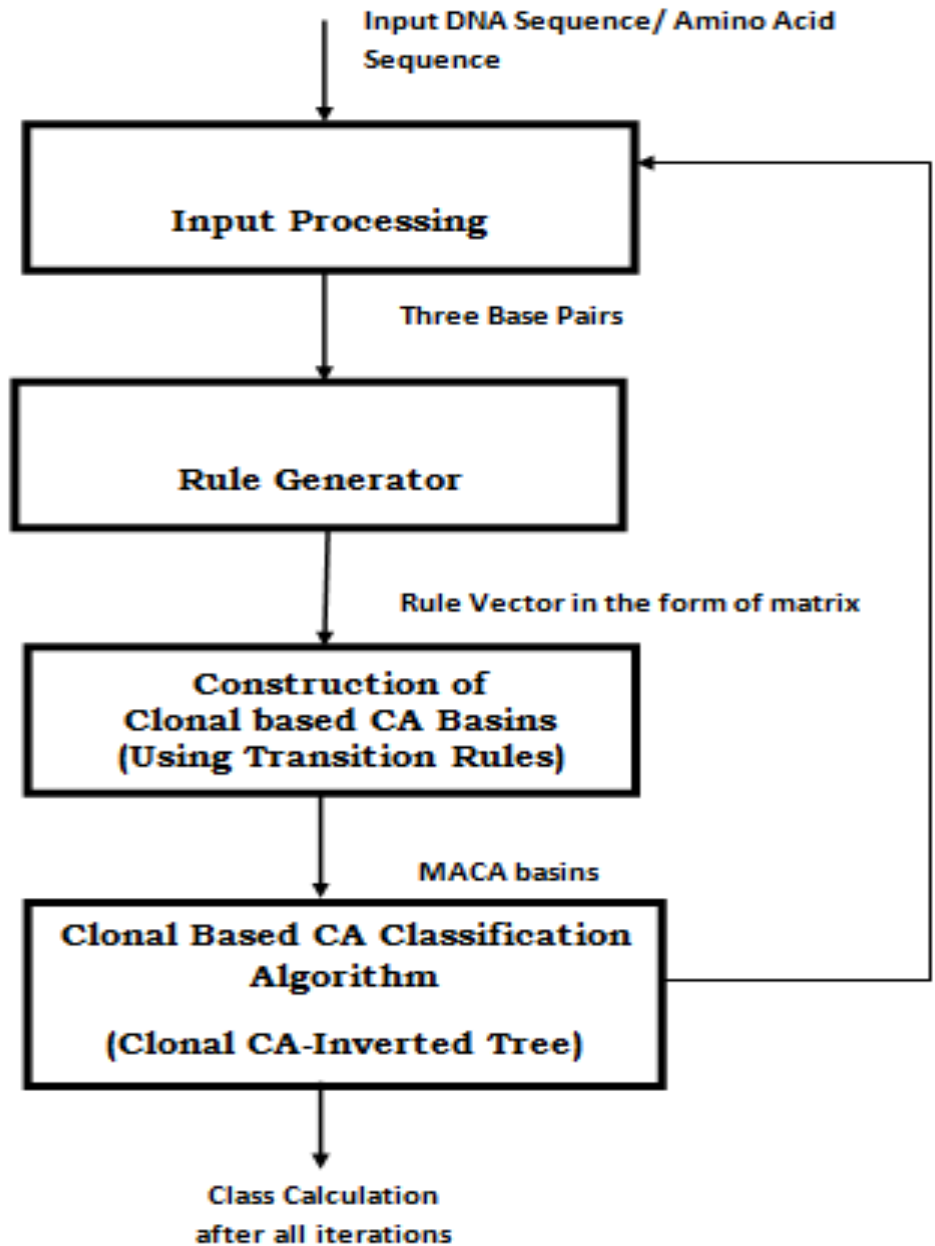

**Fig. 0:** *Design of Clonal-Based CA*

### **3.2** **Theoretical Clonal-Based CA Classifier**

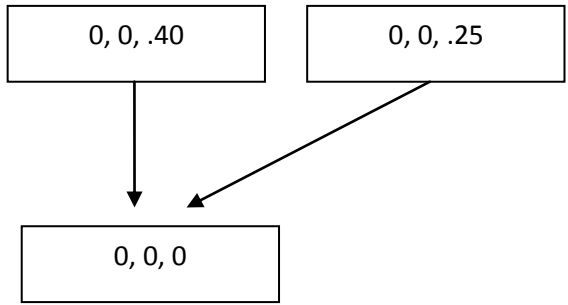

**Fig. 1:** *Basin (0,0,0)*

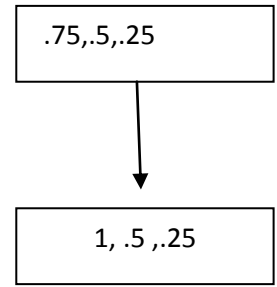

**Fig. 2:** *Basin (1,.5,.25)*

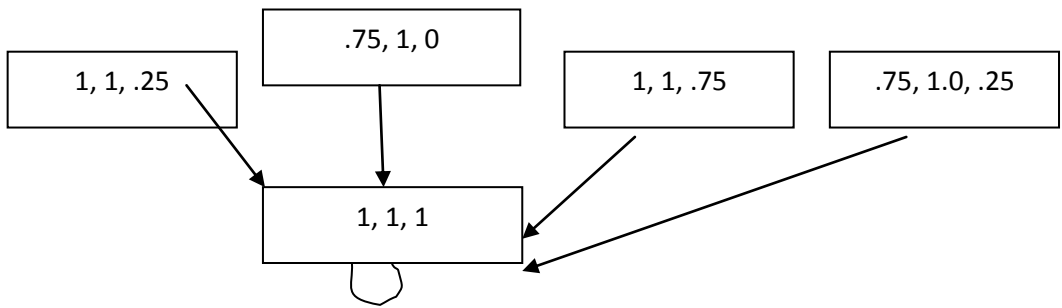

**Fig. 3:** *Basin (1,1,1)*

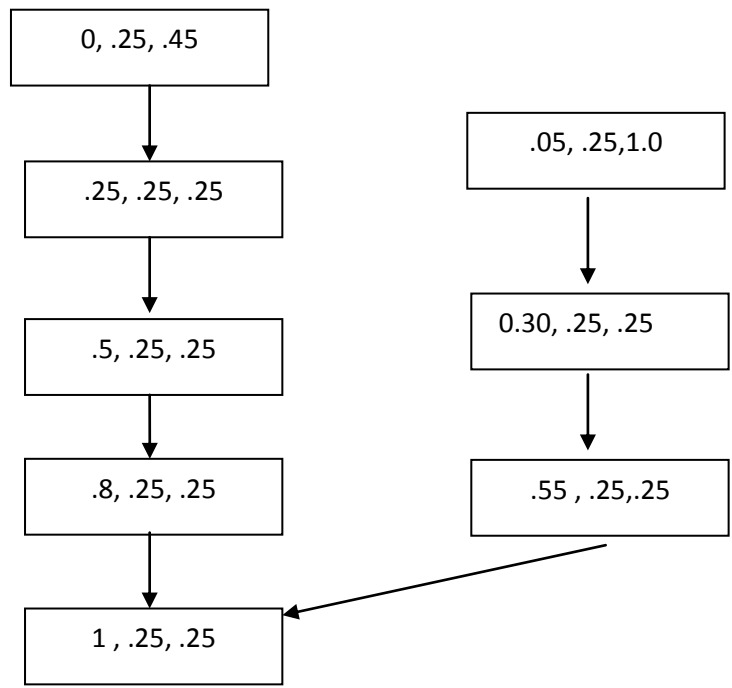

**Fig. 4:** *Basin (1,.25,.25)*

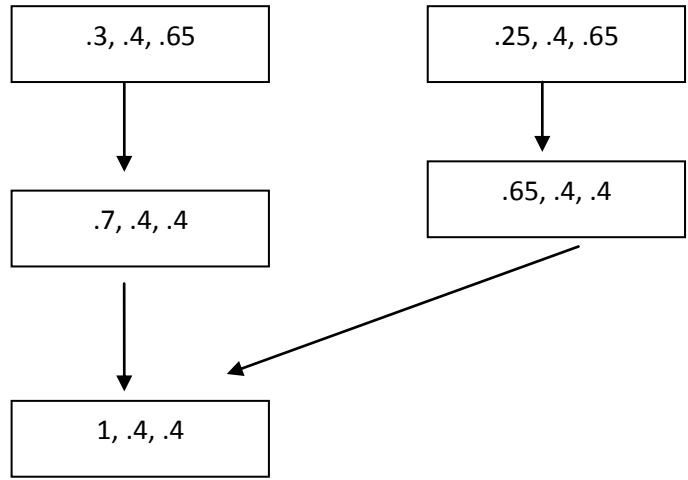

**Fig. 5:** *Basin (1,.4,.4)*

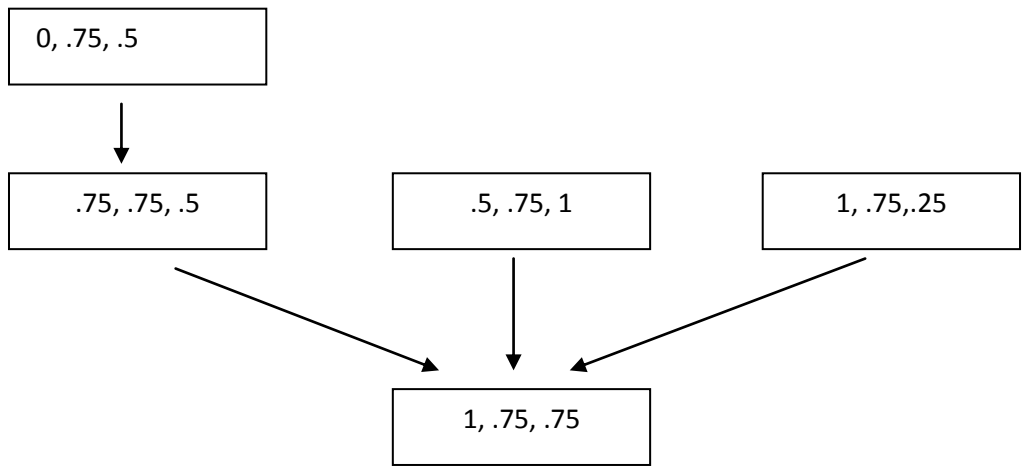

**Fig. 6:** *Basin (1,.75,.75)*

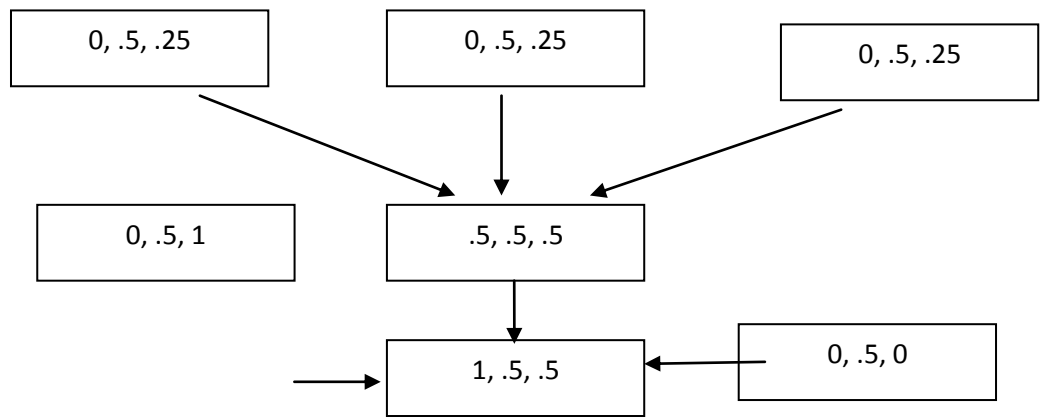

**Fig. 7:** *Basin (1,.5,.5)*

| Rule Number | AIS-FMACA Rule | Normal Rule |
|:---:|:---:|:---:|
| 254 | Min {1, $q_{i-1}(t)$+ $q_i(t)$ + $q_{i+1}(t)$ } | $q_{i-1}(t)$+ $q_i(t)$ + $q_{i+1}(t)$ |
| 252 | Min {1, $q_{i-1}(t)$ + $q_i(t)$} | $q_{i-1}(t)$ + $q_i(t)$ |
| 250 | Min {1, $q_{i-1}(t)$ + $q_{i+1}(t)$ } | $q_{i-1}(t)$ + $q_{i+1}(t)$ |
| 238 | Min {1, $q_i(t)$ + $q_{i+1}(t)$ } | $q_i(t)$ + $q_{i+1}(t)$ |

As shown in Figures 1–7, clonal-based CA assumes a real value in the range of 0–1. The rules as in Table 1 are applied and the inverted tree will be the result.

Table 1: Rules of CA

Rule< 238,250,0 > is applied to the example data set resulting in Figures 1–7.

## 4. Protein Coding Regions

Snyder[3]*et al.* have used dynamic programming with neural networks to address the protein coding region problem. Dynamic programming (DP) is connected to the issue of exactly distinguishing inside exons and introns in genomic DNA arrangements. The project Geneparser first scores the grouping of investment for graft destinations and for these intron- and exon-particular substance measures: codon utilization, nearby compositional unpredictability, 6-tuple recurrence, length appropriation and intermittent asymmetry. This data is then ordered for elucidation by DP. Geneparser utilizes the DP calculation to authorize the imperatives that introns and exons must be contiguous and non-covering and treasure the most astounding scoring combo of introns and exons subject to these requirements. Weights for different order systems are dictated via preparing a basic feed forward neural system to expand the amount of right forecasts. In a pilot concentrate, the framework has been prepared on a set of 56 human gene sections holding 150 inward exons in what added up to 158,691 bps of genomic succession. The point when tried against the preparation information, Geneparser exactly distinguishes 75% of the exons and effectively predicts 86% of coding nucleotides as coding while just 13% of non-exon bps were anticipated to be coding.

Datta[1]*et al.* used a DFT-based gene prediction for addressing this problem. Authors provided theoretical concept of three periodicity property observed in protein coding regions in genomic DNA. They proposed new criteria for classification based on traditional frequency approaches of coding regions.

Maji[11] has proposed a CA-based example classifier to recognize the coding locale of a DNA succession. CA is exceptionally straightforward, proficient, and handles more faultless classifiers than that has long ago been acquired for a reach of diverse succession lengths. Far-reaching trial outcomes show that the proposed classifier is a financially savvy elective in protein coding area identification issue.

Mena-Chalco[9]*et al.* have used modified Gabor-wavelet transform for addressing this issue.In this connection, numerous coding DNA model-free systems dependent upon the event of particular examples of nucleotides at coding areas have been proposed. Regardless, these techniques have not been totally suitable because of their reliance on an observationally predefined window length needed for a nearby dissection of a DNA locale. Authors present a strategy dependent upon a changed Gabor-wavelet transform for the ID of protein coding areas. This novel convert is tuned to examine intermittent sign parts and presents the focal point of being free of the window length. We contrasted the execution of the MGWT and different strategies by utilizing eukaryote information sets. The effects indicate that MGWT beats all evaluated model-autonomous strategies regarding ID exactness. These effects demonstrate that the wellspring of any event is some piece of the ID lapses handled by the past systems in the altered working scale. The new system stays away from this wellspring of blunders as well as makes an instrument accessible for point-by-point investigation of the nucleotide event

Changchuan Yin [17] has proposed a strategy to foresee that a protein coding area is produced which isdependent upon the way that the vast majority of exon arrangements have a 3-base periodicity, while intron groupings donot have this interesting characteristic. The technique registers the 3-base periodicity and the foundation clamor of the stepwise DNA sections of the target DNA groupings utilizing nucleotide circulations as a part of the three codon positions of the DNA successions. Exon and intron successions might be recognized from patterns of the degree of the 3-base periodicity to the foundation commotion in the DNA groupings.

# 5. Promoter Region Prediction

Gish[5] has developed an effective forecast calculation that can expand the recognition (power = 1 − false negative) of promoter. Authors introduce two strategies that utilize the machine force to ascertain all conceivable examples which are the conceivable characteristics of promoters. The primary strategy we exhibit FTSS (fixed transcriptional start site) utilizes the known TSS positions of promoter arrangements to prepare the score record that helps us in promoter forecast. The other strategy is NTSS (non-fixed TSS). The TSS positions of promoter arrangements utilized as a part of NTSS are thought to be obscure, and NTSS would not take irrefutably the positions of TSS into attention. By exploratory effects, our expectation has higher right rate than different past systems

Horwitz [6−9]has chosen an assembly of *Escherichia coli* promoters from irregular DNA groupings by swapping 19 base sets at the -35 promoter area of the Etracycline safetygene te" of the plasmid pbr322.

Substitution of 19 base setswith artificially blended irregular groupings brings about the greatest of 419 (something like $3 \times 1011$) conceivable swap groupings. From a populace in the ballpark of 1000 microscopic organisms harboring plasmids with these irregular substitutions, Tetracycline choice has uncovered numerous practical -35 promoter successions. These promoters have held just halfway homology to the -35 promoter accord grouping. In three of these promoters, the agreement operator moves 10 nucleotides downstream, permitting the RNA polymerase to distinguish an alternate Pribnow box from inside the definitive pbr322 succession. Two of the successions advertise translation more determinedly than the local promoter.

# 6. Protein Structure and RNA Prediction

Misra [10] has provided a significant part for statisticians in the period of the Human Genome Project has improved in the rising territory of structural bioinformatics. Succession dissection and structure expectation for biopolymers is a vital venture on the way to transforming recently sequenced genomic information into naturally and pharmaceutically pertinent data in backing of sub-atomic solution. We depict our work on Bayesian models for expectation of protein structure from succession, in view of dissection of a database of tentatively resolved protein structures. We have awhile ago advanced fragment-based models of protein optional structure which catches central parts of the protein collapsing procedure. These models give prescient execution at the level of the best accessible strategies. Here we indicate that this Bayesian schema is regularly summed up to fuse data dependent upon non-neighborhood arrangement cooperation's. We show this thought by displaying a basic model for strand blending and a Markov chain Monte Carlo (MCMC) calculation for induction. We apply the methodology to forecast 3-dimensional contacts for two illustration proteins.

Maji [11]worked on PSP and the arrangements submitted to the server are parsed into putative dominions and structural models are produced utilizing either near-displaying or a new structure expectation routines. Assuming that a certain match to a protein of known structure is found utilizing BLAST, PSI-BLAST, Ffas03 or 3d-Jury, it is utilized as a pattern for relative displaying. Assuming that no match is found, structure expectations are made utilizing again

Rosetta piece insertion strategy. Test atomic attractive thunder (NMR) obligations information can likewise be submitted with an inquiry succession for Rosetta NMR all over again for structure determination. Other current capacities might be the expectation of the impacts of transformations on protein–protein communications utilizing computational interface alanine filtering. The Rosetta protein configuration and protein–protein docking systems will soon be accessible through the server

Pokkuluri [12] predicts protein tertiary structure on the premise of amino harsh corrosive succession remains one of the extraordinary issues in biophysical science. As per the thermodynamic speculation, the local compliance of a protein could be anticipated as the worldwide best of its free vigor surface with stochastic streamlining techniques requests of size quicker than by immediate reenactment of the collapsing procedure.

## 7. Clonal-based CA

The proposed tool(Figures8–10)is tested for predicting the protein coding and promoter region using the same classifier. We then extended this framework to predict the secondary structure of the protein. The classifier is trained with Fickett and Toung data sets for protein coding regions and *E.Coli* data sets for promoter prediction. This proposed algorithm can handle different sequence lengths input and can give an accuracy of 82%.

*4*

| Promoter Sequence: | | |
|---|---|---|
| 1 | AAAAAAAAAA AAAAAAAAAA AGGGGGGGGG GGGGGGGGGG GGGGGGGGGG GGGGGGGGGG GCCCCCCCCC CCCCCCCCCC | 80 |
| 81 | CCCCCCCCCC CCCCCCCCCC AGCAGCAGCG GGGGGGGGGG GGGGGGGGGG GGGGATAAAA AAAAAAAAAA AAAAAAAAAA | 160 |
| 161 | TGATAGAAAA AAAAAAAAAA AAAAAAAAAA AAAAAAAAAA AAAAAAAAAA TAAAAAAAAA AAAAAAAAAA AAAAAAAAAA | 240 |
| 241 | AGCAGCAAAA AAAAAAATAA AAAGCTGCAA AAAAAAAAAA AAAAAAAAAA AAAAAAAAAA AATAAAAAAA AAAAAAAAAA | 320 |
| 321 | AAAAAAAAAA AAAAAA | |

***Fig. 8:*** *Common Protein Coding and Promoter Prediction Interface*

| Start Position | End Position | Strand | Score | Sequence |
|---|---|---|---|---|
| 298 | 313 | - | 23.06 | TTTTTTTTTTATTTTT |
| 206 | 221 | - | 23.06 | TTTTTTTTTTATTTTT |
| 131 | 146 | - | 21.43 | TTTTTTTTTTATCCCC |

***Fig. 9:*** *Boundary Reporting*

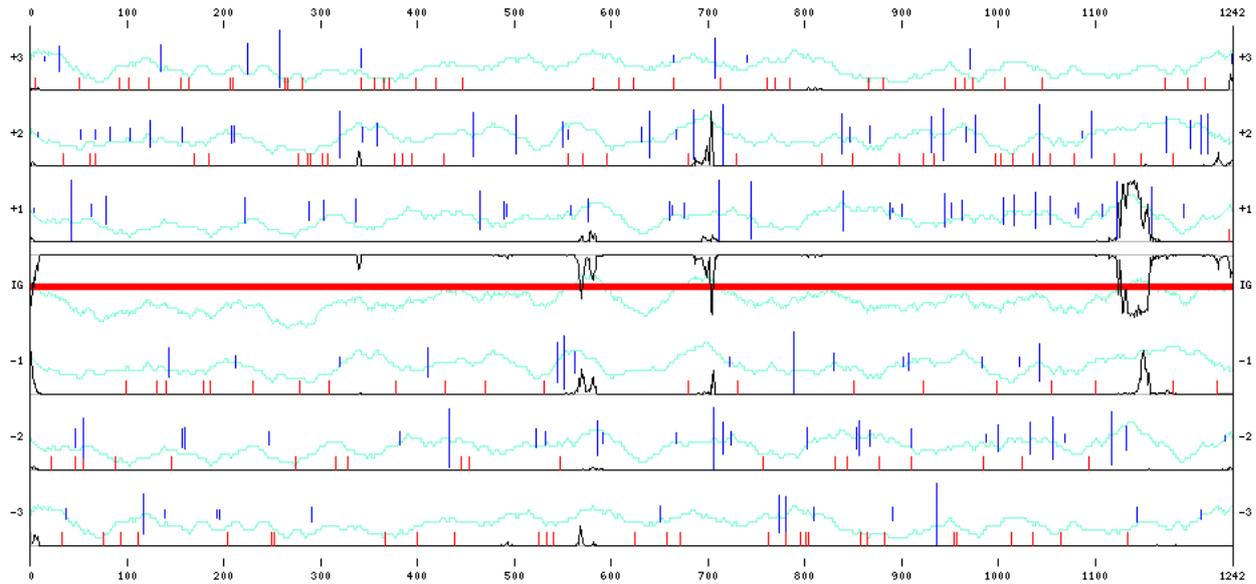

***Fig. 10:*** *Protein Coding Region Prediction*

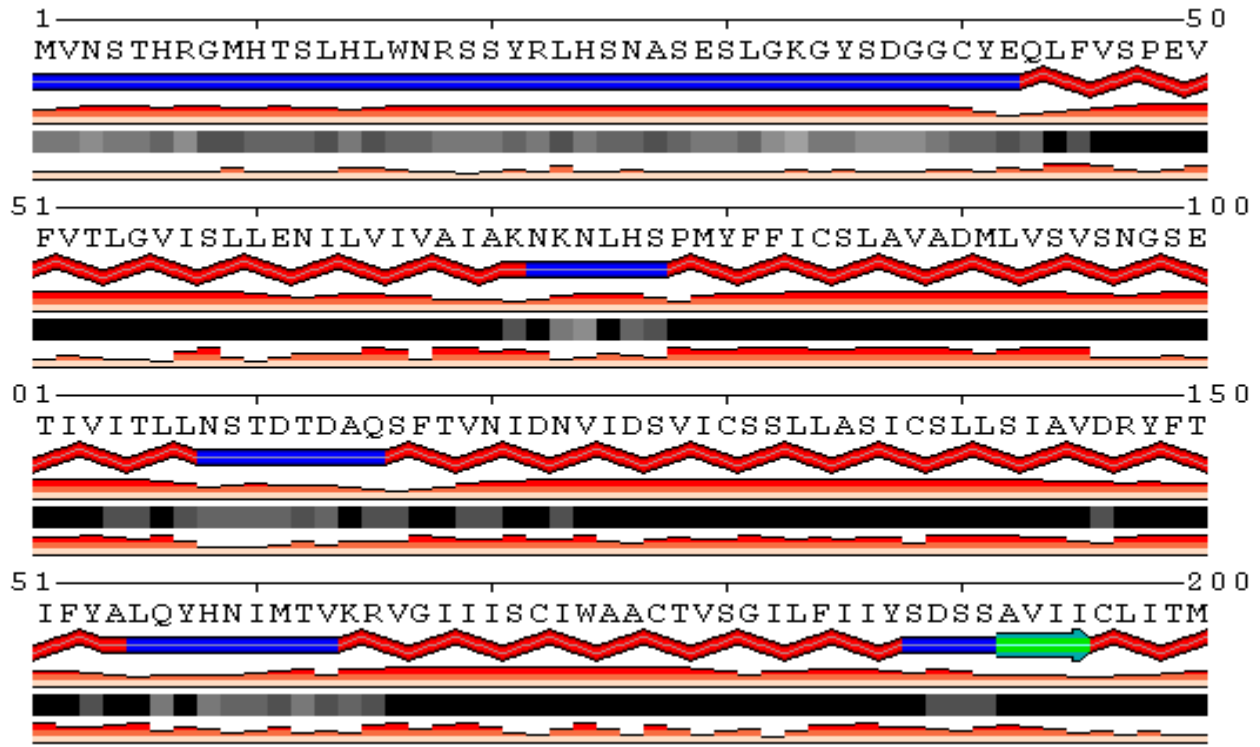

***Fig. 11:*** *RNA and Protein Structure Prediction Interface*

## 8. Conclusion

This paper is aimed at bringing an intuition towards application of CA in bioinformatics. A common framework is required for addressing major problems in bioinformatics.CA framework can be used to address many problems like protein structure prediction, RNA structure prediction, predicting the splicing pattern of any primary transcript and analysis of information content in DNA, RNA, protein sequences and structure and many more. This work can be extended to address many other problems in bioinformatics.